\PassOptionsToPackage{hyphens}{url}
\documentclass[letterpaper,twocolumn,10pt]{article}
\usepackage{usenix}
\usepackage{authblk}
\usepackage{epsfig,endnotes}
\usepackage{cite}
\usepackage{amsmath,amssymb,amsfonts}
\usepackage[ruled,linesnumbered]{algorithm2e}
\usepackage{multirow}
\usepackage{makecell}
\usepackage{algorithmic}
\usepackage{dblfloatfix}

\usepackage{subcaption}
\usepackage{pifont}
\usepackage{comment}
\usepackage{graphicx}
\usepackage{textcomp}
\usepackage{xcolor}
\usepackage{hyperref}
\usepackage{array} 
\usepackage{booktabs}
\usepackage{enumitem}
\usepackage{placeins}
\usepackage{makecell}
\usepackage{caption}

\SetKwComment{tcp}{$\triangleright$ }{}

\newcommand{\xuehai}[1]{\textcolor{red}{#1}}
\newcommand{\yuhao}[1]{\textcolor{black}{#1}}
\newcommand{\yuhaosuggest}[1]{\textcolor{black}{#1}}

\begin{document}

\date{}

\title{\Large \bf Redox: Improving I/O Efficiency of Model Training Through File Redirection}
\author[1]{Yuhao Li}
\author[1]{Xuanhua Shi}
\author[1]{Yunfei Zhao}
\author[2]{Yongluan Zhou}
\author[1]{Yusheng Hua}
\author[3]{Xuehai Qian}
\affil[1]{Huazhong University of Science and Technology}
\affil[2]{University of Copenhagen}
\affil[3]{Tsinghua University}

\maketitle 



\subsection*{Abstract}

This paper proposes Redox, a training data management system designed to achieve high I/O efficiency.
The key insight is a new observation of {\em file redirection}: 
for model training, when  training data in one file is requested, 
the system has the flexibility
to return the data of another file. 
Based on this property, 
Redox starts with a bold design principle that chunks of 
data files are always read from disk in batch, and
once loaded, all files in the chunk will be consumed without 
being loaded again. 
We propose efficient local and distributed file read protocol
based on this principle that both minimizes the wasted data read
and enables opportunistic prefetch from remote node. 
Moreover, we analyze file redirection's impact on 
randomness, and show that it has little effects
on training efficiency. 
Experimental results indicate that Redox significantly accelerates data fetching in training, achieving up to a 4.57x improvement in end-to-end training compared to PyTorch. 

\section{Introduction}

\label{sec:introduction}
 Deep learning ~\cite{lecun2015deeplearning,schmidhuber2015deepneuralnetworks} has permeated various domains, including computer vision~\cite{krizhevsky2012alexnet,he2016resnet,howard2017mobilenets}, speech recognition~\cite{hinton2012speech,graves2013speech,amodei2016speech}, and natural language processing~\cite{mikolov2013nlp,vaswani2017attentionnlp,devlin2018bertnlp}, becoming ubiquitous in society. The widespread adoption of deep learning can be largely attributed to its remarkable accuracy, a quality dependent on not only deep learning model structures such 
  as Deep Neural Network (DNN) or transformer 
  but also the training datasets. 
 Large amount of high quality training datasets with rich
 diversity are crucial in achieving high accuracy of machine learning (ML) models. 

 Due to scaling law~\cite{kaplan2020scaling}, the size of training datasets continuously grows. 
 For instance, ImageNet-21k~\cite{deng2009imagenet} spans 1.1TB with tens of millions of images, while modern multimodal datasets such as MINT-1T~\cite{awadalla2024mint} (over 300TB), LAION-5B~\cite{schuhmann2022laion} (~300TB), and COMMONPOOL~\cite{gadre2023datacomp} (~450TB) contain billions of image–text pairs.
 The massive scale of datasets imposes significant I/O pressure on ML model training. 
 In particular, the current trend of improving efficiency
 and performance of computation through hardware, e.g.,
 GPUs, NPUs and TPUs~\cite{}, and
 system innovations will further exaggerate the severity of
 I/O bottleneck since the gap between the processing speed and 
 moving data out from disks~\cite{mohan2021coordl,khan2023shade,kumar2020quiver}
 is further increased. 

\begin{figure*}[htbp]
    \centering
  \vspace{-5mm}

    \includegraphics[width=\linewidth]{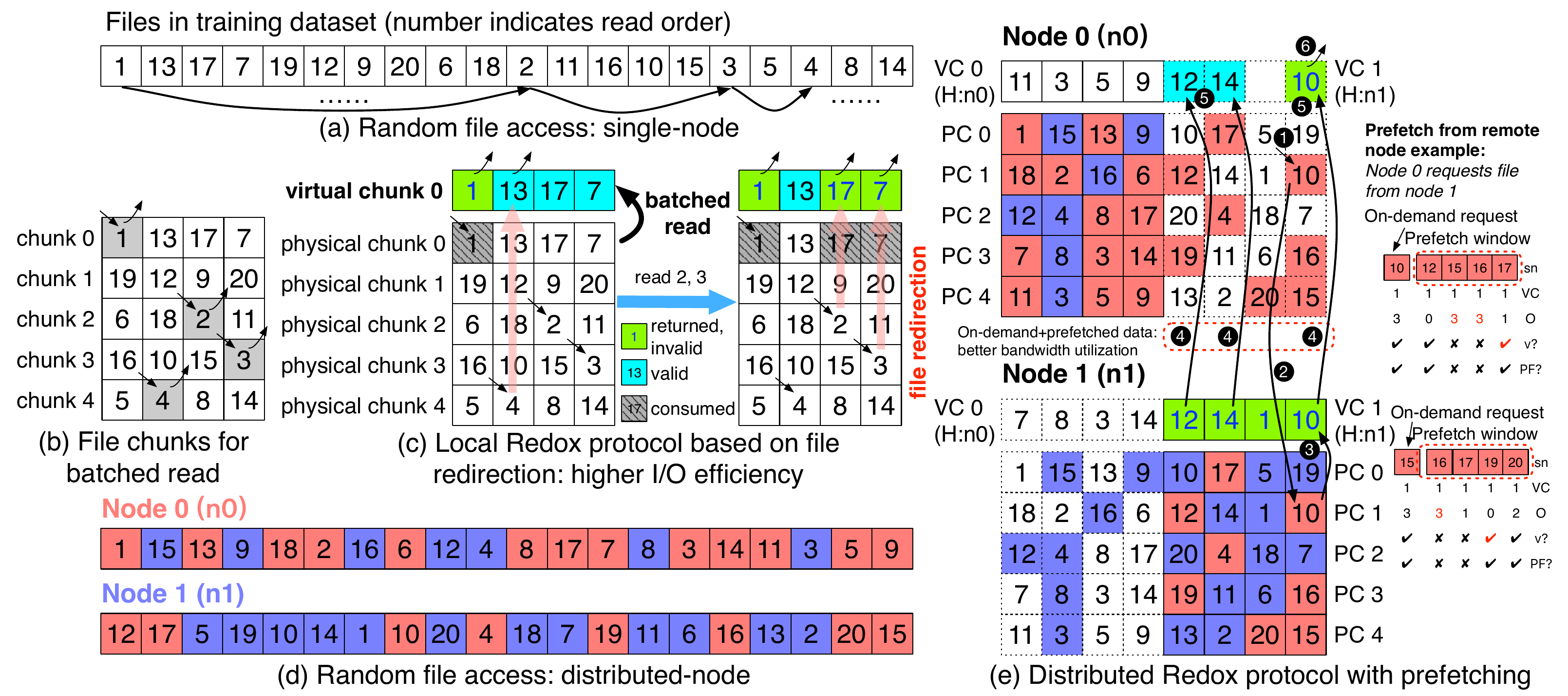}
  \vspace{-7mm}

  \caption{File Redirection: Key Insights of Redox Protocols
  }
  \vspace{-6mm}
  \label{fig:overview}
\end{figure*}

The I/O bottleneck due to the increasing amount of training 
data is difficult to tackle because of
its random access pattern shown in Figure~\ref{fig:overview} (a). 
In general, the efficiency of storage devices (e.g., SSD)
is higher with sequential read
e.g., a high-end NVMe Gen4 SSD such as the Samsung 980 Pro~\cite{Samsung980Pro} achieves up to 7,000 MB/s in sequential reads, but only about 1,000K IOPS in random 4K reads—equivalent to just ~4,100 MB/s effective throughput.
The random access pattern makes the batched file read shown 
in Figure~\ref{fig:overview} (b) ineffective and 
prevents the 
techniques that reorder disk accesses to 
enforce sequential read.
Another common technique is to use cache to keep
frequently accessed data in faster storage.
Besides the random access, which prevents the exploration of spatial locality, the ``read once'' 
property makes temporal locality useless: 
each file is read and consumed to train
the model only once.
Most existing methods~\cite{mohan2021coordl, yang2019locality-aware, dryden2021nopfs} cache
a fixed portion of data in memory, enabling faster
access for the cached data while incurring
high I/O latency for the uncached data. 
Based on such static caching policy, 
the proportion of data access that can be 
accelerated depends on the memory capacity for 
caching data, which, unfortunately, 
does not scale with training data size.

This paper rethinks file access during
ML model training and 
proposes a novel technique that achieves the {\em batched
sequential file read} while mostly {\em preserving the randomness}
that ensures the training efficiency.
The {\bf crux} of our technique is a new observation for ML training:
when the training data in file A is requested, 
the system has the flexibility
to {\em return the data of another file B}. 
The seemingly surprising observation is unique for 
ML training, of which the whole purpose is to go over all files in training dataset
in some random order during an epoch.
When a particular file is requested from the training process,
what really needed is the data of {\em some random file} in training dataset. 
For the first time, this paper reveals the unique opportunity we name as {\em file redirection}
and capitalizes on it to significantly improve training I/O efficiency.

Building on the surprising property of file redirection,
we propose {\em Redox}, a new training data management system designed to achieve high I/O efficiency.
It partitions all files of the training dataset into {\em chunks}, each of that
 contains a predefined number (chunk size) 
 of consecutively stored files in disk of a training node, named as {\em physical chunks}.
 In memory, each training node maintains a fixed number of
 {\em virtual chunks} with the same chunk size as physical chunks.
Similar to the mapping from memory block to cache block in 
a conventional cache system, physical chunks are mapped
to virtual chunks using a certain mapping function. Multiple
physical chunks can be mapped to the same virtual chunk, since 
the number of virtual chunks is fixed and smaller than the 
total number of physical chunks.
Figure~\ref{fig:overview} (c) shows an example of virtual and physical
chunk, and file redirection, it will be explained in detail
later in Section~\ref{sec:examples}.
This organization works for both single-node and distributed-node setting. Each physical chunk has a ``home'' node
where the file is actually stored. The virtual chunks can be classified 
into different types: {\em local virtual chunk}, which can be mapped
with local physical chunks; and {\em remote node $i$ virtual chunk},
which can 
be mapped with physical chunks stored in node $i$. 
While seemingly resembling a typical cache system, 
Redox diverges into an entirely
new system leveraging the flexibility of file redirection.

We first outline the single-node setting to introduce the 
key features. Redox starts from two aggressive policies:
(1) all file reads from disk
are {\em batched and sequential}; and 
(2) each file is loaded into memory {\em exactly once}. 
When a file A is requested, assuming
that the virtual chunk for A is empty, 
the whole physical chunk containing A is read from disk and 
filled into the corresponding virtual chunk,
then the data of file A is returned and file A's
slot in the virtual chunk is {\em self-invalidated} 
due to (2).
Also based on (2), all files in the filled virtual chunk will be {\em consumed}
by training process without being evicted to disk and being loaded again,
even if the {\em on-demand} request, i.e., the request
that triggers the loading of its physical chunk to 
the target virtual chunk, takes just one of
the slots in virtual chunk.
A ``consumed'' bit is 
associated with each file to ensure (2): once loaded to virtual chunk,
``consumed'' bits of all files in the physical chunk
are set.

When a different file B is requested, if its slot in the target virtual chunk is non-empty but 
does not contain file B, it means that the virtual chunk
contains the data of a different physical chunk (e.g., file C's 
physical chunk)
mapped to the same virtual chunk.
In conventional cache, a cache block eviction is required, and the
block containing requested data will be refilled into the cache.
In Redox, thanks to file redistribution, as long as the slot for file B in
the current target virtual chunk contains valid data, this in-memory data
can be returned as the response for file B request, saving one disk access. 

However, if the current virtual chunk is non-empty, but the slot
for file B is invalid, the system has no choice but filling the slot
with data from disk. Due to policy (2), the {\em valid but un-consumed}
data in the virtual chunk cannot be evicted, but due to (1), an 
entire different physical chunk will be read from the disk and replace
current data of the virtual chunk. To reconcile the conflicting 
requirements, Redox has to incur {\em I/O waste} by reading the
whole physical chunk but not replacing the valid but un-consumed
files in the current virtual chunk.
For those files which are read from disk
but not filled into memory, the ``consumed'' bit is not set 
after the read, which means that they may be read again. 
Note that it does not conflict with policy (2), since they are not
loaded to memory. Naturally, when the system needs to 
(partially) refill a virtual chunk, it needs to make {\em correct and
good} decision in selecting a physical chunk to read.
On the correctness side, file B's corresponding slot in the candidate 
physical must be not consumed; on the optimization side, 
the wasted read should be minimized by choosing the physical
chunk that converts more files to ``consumed'' state.
Both considerations are incorporated in the complete Redox protocol.

The distributed-node case is a natural extension to single-node
mechanism. 
In Redox, each node has a view of the 
same number of virtual chunks (local + remote). 
For example, for a system with 4 nodes and 16 virtual chunks in total, 
each node logically
has access to 4 local virtual chunks
and 12 remote virtual chunks (4 for each remote node).
The remote data access procedure is invoked transparently
based on the file requested which will be statically
mapped to one of the remote (e.g., node B) virtual chunk
and be served by node B.
In the local protocol, data is always read from disk 
in batch, but returned to the requester the individual file's data.
When node A requests the file data stored in remote node B, node B leverages the same local data access protocol to read data from disk in batch if necessary,
then, sends the data of the requested file to node A.
For a given node, in distributed setting,
the consumers of file data not only include the local
training process but also all nodes in the system, 
thus, data in virtual chunk will be consumed faster, 
which may impact overall performance.    
Nevertheless, functionally, the remote protocol can work
seamlessly with the local protocol.

The performance of inter-node
communication is lower and may become the bottleneck. 
To this end, the basic distributed protocol can be 
enhanced
with {\em opportunistic prefetching}. The basic idea is to prefetch and transfer additional data 
(if already available in memory) 
for the requester's future reads together with the 
on-demand file, achieving better utilization of communication
bandwidth. To realize it, three problems need to be considered:
(1) what to prefetch?
(2) how to prefetch?
(3) how to correctly prefetch?
Different from traditional applications with certain access patterns
that can serve as the base for predictive prefetch, the file access
for training is completely random. To this end, we propose to replicate and distribute the {\em predetermined} file access traces of each node to
all other nodes, so that when a node A receives an on-demand request from node B, node A can accurately get the next few file
accesses of node B. Then, the prefetch can specifically target these files, 
without introducing any waste. 
With the accurate prefetch list, one choice is to initiate them 
earlier, however, doing so will unnecessarily
increase the data movement
between disk and memory. Thus, we took an opportunistic
approach: only prefetch the data that is already in a virtual chunk. 
A side-effect of prefetching is that virtual chunk slots
will become empty more quickly, which may reduce the data
waste due to the un-consumed file slot. 

The correctness requirement of prefetching is: the prefetched files 
sent from remote node to the requester node 
cannot be mapped to the same slot of the same virtual chunk. 
It is because only one file can be inserted to the slot, 
for others, there
is no buffer to keep the data. 
To ensure this property, when an on-demand request is
received, the remote node considers a file access list starting with 
the on-demand request 
in the trace of the requester. The node can scan the list based
on the access order starting from the on-demand request, generating
a pair of (virtual chunk, offset) for each request. If two requests
are mapped to the same pair, then this access is skipped.
Figure~\ref{fig:overview} (d) and (e) show examples of 
remote read with prefetching, see details in 
Section~\ref{sec:examples}.

We implement an open-source prototype of Redox~\footnote{\url{https://github.com/Redox-Project/Redox}} and integrate it with PyTorch. We conducted extensive experiments using three datasets, multiple deep neural network (DNN) models, and three different hardware setups representing scenarios with less and more computing power, respectively. The results indicate that, while ensuring the convergence of model training, Redox consistently outperforms vanilla PyTorch by up to 4.57 times, with peaks reaching 1.96 times speedup that of related work, particularly under limited memory constraints.

\section{Background}
\label{sec:background}
\subsection{DLT's I/O Bottleneck}

When memory capacity is insufficient to cache the training dataset in memory, I/O becomes the performance bottleneck for model training,
the random data access pattern further exaggerates the 
challenge. 
To validate the bottleneck, we perform experiments to measure the I/O overhead of training different models on the ImageNet-1k dataset (135 GB)~\cite{deng2009imagenet}. The experiments are conducted on three nodes, each with one NVIDIA Tesla P100 GPU and a memory capacity of 60 GB. We use ``epoch time'' to refer to the time taken to complete one epoch of end-to-end training and ``I/O overhead percentage'' to refer to the percentage of epoch time solely occupied by I/O overhead. The I/O overhead is measured by comparing training time with data read from storage and training time with data dynamically generated by a data generator without any I/O. Further information on the experimental setup can be found in Section~\ref{sec:evaluation}.

As shown in Table \ref{tab:overheadOfDataLoading},  the experimental results indicate that in the training of all models, I/O overhead occupies a significant percentage of epoch time, ranging from 65\% to 91\%.
In the training of ResNet50~\cite{he2016resnet}, a model with heavier computation, the I/O overhead still accounts for 65\% of the training time. For lightweight models such as SqueezeNet~\cite{iandola2016squeezenet}, the I/O overhead is up to 91\%. The I/O overhead can be attributed to the following reasons.

\begin{table}[ht]
  \centering
  \vspace{-3mm}
  \caption{ I/O Overhead in DLT (hours)}
    \vspace{-3mm}

    \label{tab:overheadOfDataLoading}
    \begin{tabular}{
        >{\centering\arraybackslash}m{0.2\linewidth}
        >{\centering\arraybackslash}m{0.1\linewidth}
        >{\centering\arraybackslash}m{0.2\linewidth}
        >{\centering\arraybackslash}m{0.2\linewidth}
    }
    \toprule
    DNN Models  & Epoch Time & I/O Overhead Time&  I/O Overhead Percentage\\
    \midrule
     SqueezeNet  & 1.40  & 1.27  & 91\%\\
    MobileNetV3 & 1.53 & 1.25 &82\%\\
    ResNet50 & 1.65 & 1.07 & 65\%\\
    \bottomrule
    \end{tabular}
  \vspace{-3mm}

\end{table}

In model training, the gradient descent algorithm (e.g.,  \textit{stochastic
gradient-descent} (SGD)~\cite{bottou2018sgd}, Adam~\cite{kingma2014adam}, Momentum
~\cite{yu1993Momentum}) is typically run for many epochs to adjust the model's weights and parameters until the model converges. In each epoch, the algorithm goes through the entire dataset with certain random order. To realize this, popular deep learning frameworks like PyTorch shuffle the dataset indices at the start of each epoch using a random seed, creating a random access sequence. 
Then the frameworks access data following this random access sequence in this epoch. Each data item will be accessed once at a unpredictable time during each epoch. 
Due to the random access pattern, caching 
mechanism with classical replacement strategy is ineffective
in improving performance. 



Furthermore, inputs to the training process are often stored in many small files, which incur higher I/O overhead than large files when 
randomly accessed.
To minimize the overhead, large-scale analytics systems~\cite{ghemawat2003gfs,shvachko2010hadoop} typically re-organize them into relatively large data chunks and employ batched reading. However, due to the random access pattern, directly
applying batched reading to model training is not effective. 
Specifically, when a chunk of data is loaded from 
disk, there is no way to ensure that
all or most of the data contained in the chunk are
consumed before being swapped
out of memory.
We believe that file redirection is the key idea
that can fill this gap. 

\subsection{Related Work}


DIESEL+~\cite{wang2021diesel+} aims to alleviate training's I/O bottleneck of reading 
small files by batched reading and grouping training data into data chunks to improve performance. To support random data access, shuffling is performed at the chunk
level. 
Compared to Redox, DIESEL affects randomness more. 
Using the chunk abstraction of Redox, DIESEL's policy
essentially forces the files in a chunk to be accessed 
consecutively, while Redox largely preserves the original randomness
at file level. We will discuss the effects on randomness
in more detail in Section~\ref{}.
Some approaches~\cite{mohan2021coordl, yang2019locality-aware, dryden2021nopfs} aim to optimize memory management by caching a fixed portion of data in memory without eviction. These approaches guarantee that the data cached in memory will be hit once per epoch.
This approach is inherently limited by 
memory capacity. 

Some other approaches~\cite{2018zhudeepio, 2022nguyenPartial-Shuffle} focus on optimizing I/O for distributed training by localized shuffling. They allow each node to independently shuffle locally cached or stored data to alleviate low memory hit rates caused by global shuffling. Additionally, they employ techniques like double buffering and cross-node data sharing to overlap disk I/O.
 However, these approaches compromise global randomness, 
  impacting model convergence as indicated by their experiments.

SHADE~\cite{khan2023shade} presents a novel method for generating access sequences based on data importance scores. It suggests that not all data have equal importance for model convergence, advocating for training high-importance data multiple times within a single epoch and caching them in memory to enhance hit rate and training performance. This approach deviates from the conventional practice of training on each data item once per epoch. While this method may work well in its targeted scenario, 
its generalizability is unknown. 
Nevertheless, our proposed mechanisms can still accelerate
the I/O performance for the data not belonging to 
the high-importance set. 




\subsection{Is I/O Truly Important?}

The focus of this paper and related work is 
to better utilize the data already loaded into memory, 
one natural question is: why not equipping the system with 
sufficient memory large enough to accommodate the entire 
training dataset? We argue that it is not viable based on the 
current trend. 

On the one hand, the size of training datasets to achieve high 
accuracy is continuously growing. 
For example, public datasets have grown from 1.1TB for ImageNet~\cite{deng2009imagenet} to 
450TB for COMMONPOOL~\cite{gadre2023datacomp} 
in just over a decade.
On the other hand, the cost of memory to store such large
amount of data is far too high for most users, preventing 
this approach to be adopted as a common solution. 
For example, to store 450TB training data, the cost
	of memory can reach three million US dollars. 
    To store the same amount of data
	in distributed nodes, with a reasonable memory size of 256GB each
	node, 1800 nodes are required. To enable distributed 
	training, each node needs to have at least one GPU, this leads
	to 1800 GPUs and a cost of 27 million USD. 
    Moreover, as the number of nodes
	increases, the communication overhead also grows considerably.
	If we use a smaller number of nodes with larger memory in each,
	the cost of memory will dominate again. 
Overall, keeping the entire training dataset in memory is 
impractical, incurring high hardware cost.

\section{Redox: Caching System with File Redirection}
\label{sec:design}

This section discusses detailed design of
Redox, a new training data management system built
based on the idea of file redirection. 
We first describe the system overview, then define the 
key system components as abstract system objects with
various attributes.
Based on the abstract system components, we specify the local
and distributed protocol for file data access that cleanly 
interfaced with training process.
Finally, we explain two running examples to provide the
concrete understanding.

\subsection{Redox Overview}

The Redox system is a drop-in middle layer between ML training framework such as PyTorch and the underlying hardware
resources local memory and disk. It also handles the communication with remote nodes.
Before the execution of each epoch, 
training frameworks can generate a new random data access sequence as normal and request data from Redox according to the access sequence.
Redox would return file data based on Redox protocol, 
with the key distinction that the returned file may not be exactly
the one requested according to the access sequence. 
Functionally, Redox returns individual file data from memory, orchestrates
batched read from disk, sends remote file read request and processes the 
responded data from remote node. 

Redox is composed of two logical layers:
(1) the mapping layer handles file-to-memory, memory-to-chunk and physical-to-virtual chunk mapping; and
(2) the data layer is responsible for local data access, chunk management, and communication
with remote nodes. 
During execution, the mapping layer redirects data requests from the training framework to the internal memory locations before initiating batched data reading if the 
target virtual chunk does not contain valid data in the file's slot.
The data layer will select the proper physical chunk to be filled into memory. 

\subsection{System Parameters}
\label{sec:sys_parameter}

\begin{table}[h]
	\begin{center}
		
		\footnotesize
        \vspace{-5mm}
		\begin{tabular}{|p{1cm}||p{6.4cm}|}
			\hline 
			Name &  Explanation \\
			\hline
			F & Total number of files across all training nodes  \\
			\hline
			M & Total number of virtual chunks across all training nodes\\
			\hline
			K & Chunk size: number of files in a virtual or physical chunk\\
			\hline
		    N & Total number of training nodes \\
			\hline
			PCS size & \yuhaosuggest{$F/(KM)$: \# of physical chunks mapped to a virtual chunk} \\
            \hline
            P & The size of prefetch buffer \\
			\hline
		\end{tabular}
	\end{center}
    \vspace{-6mm}
	\caption{System Parameters}
	\vspace{-4mm}
	\label{table:sys_config}
	\end{table}

Table~\ref{table:sys_config} provides a few key parameters
on the sizes of various data components. Initially, all $F$ files are partitioned and
stored in the disk across all nodes. 
In each node, files are stored consecutively but accessed randomly. 
$M$ is the number of virtual chunks (VCs)
across all nodes and all nodes have the same view of VCs. But for a given VC,
only one will hold the data from its local files, for other VCs in other nodes, 
file data will be requested the ``home'' node and filled in. 

$K$ is a system-wide
parameter for the virtual and physical chunk size. 
Determining chunk size requires the consideration 
of a tradeoff between throughput and execution time. 
A larger chunk size generally reduces the I/O overhead but on the other side, it would increase the number of times a chunk is loaded within an epoch, leading to more data transfer waste.
The good chunk size choice can be obtained through 
an one-time profiling for each training platform.
 In our system, we choose 64 as the optimal choice 
based on the results reported in Section~\ref{sec:eveluation-chunkSize}. 
Given $K$, the number of physical chunks (PCs) across all nodes 
is $F/K$, thus, each VC is associated with a physical chunk set (PCS) of 
size $F/(KM)$.
The relationship between VC and PCS is similar to a cache block and a multi-way
cache set in a typical memory hierarchy.
Note that physical chunk generation is an one-time process, and the pre-organized data chunks can be re-used to train different models with different random
access orders in different epochs.
The mapping between a virtual and the physical chunks in PCS is stored in table 
generated once and accessed by the mapping layer. 
Although we specify a constant $M$ of VCs with constant size $K$,
the memory for each VC is {\em variable} depending on the file sizes of the mapped PC
and is dynamically allocated. 

\subsection{Abstract Data Organization}
\label{sec:abs_data}

\begin{table}[h]
	\begin{center}
\vspace{-4mm}		
		\footnotesize
		\begin{tabular}{|p{1.4cm}||p{2.1cm}|p{3.6cm}|}
			\hline 
		   Global Data & Value Range or \newline Index Range & Explanation \\
			\hline
			{\bf \em File[F]} & File[0:(F-1)]  & All files in training dataset \\
			\hline
			\thickspace \thickspace {VC} & $[0,...,(M-1)]$ & The file's virtual chunk ID \\ 
			\thickspace \thickspace {PC} & $[0,...,(F/K-1)]$ & The file's physical chunk ID \\ 
			\thickspace \thickspace {O} & $[0,...,(K-1)]$ & Offset inside the chunk \\
			\thickspace \thickspace {H} & $[0,...,(N-1)]$ & Home node of the file \\
			\thickspace \thickspace {R} & $[0,...,(N-1)]$ & Requester node of the file \\
			\thickspace \thickspace {sn} &  $[0,...,(F-1)]$ & The sequence number of the file \\    
            \thickspace \thickspace {addr} &   & The starting address of file data \\   
            \thickspace \thickspace {consumed} &true or false & The file has been consumed? \\ 			
            \hline
			{\bf \em VC[M]} & VC[0:(M-1)]  & A virtual chunk across all nodes \\
			\hline
            \thickspace \thickspace {VFS[K]} & VFS[0:(K-1)] & VC file slot \\ 			
             \thickspace \thickspace {VFS\_v[K]} & VFS\_v[0:(K-1)] & VC file slot contains valid file? \\
             \thickspace \thickspace {addr\_v[K]} & addr\_v[0:(K-1)] & The VC slot's file address \\         
            \thickspace \thickspace {\yuhaosuggest{PCS[$\frac{F}{KM}$]}} & PCS[0:($\frac{F}{KM}-1$)] & Physical chunk set\\ 			
            \thickspace \thickspace {H} & [0,...,(N-1)] & Home node of the virtual chunk \\ 			
            \hline
            {\bf \em PC[F/K]} & PC[0:(F/K-1)]  & A physical chunk \\
            \hline
            \thickspace \thickspace {PFS[K]} & PFS[0:(K-1)] & Physical chunk file slot  \\ 			
            \thickspace \thickspace {VC} & $[0,...,(M-1)]$ & The VC this PC is mapped to \\ 			
            \thickspace \thickspace {H} &[0,...,(N-1)] & Home node of the PC \\ 
            \hline
            {\bf \em PFB[N]} & PFB[0:(N-1)] & Prefetch buffer for each node \\
            \hline 
            \thickspace \thickspace {map[N][P]} & map[0:(N-1)][0:(P-1)] & Bitmap of prefetched in the window from an on-demand request\\ 
            \thickspace \thickspace {p\_VC[N][P]} & p\_VC[0:(N-1)][0:(P-1)] & VC idx. of prefetched in the window from an on-demand request\\ 
            \thickspace \thickspace {p\_O[N][P]} & p\_O[0:(N-1)][0:(P-1)] & Offset of prefetched in the window from an on-demand request\\ 
            \thickspace \thickspace {sn[N]} & sn[0:(N-1)] & Sequence number of current on-demand request
            \\ 
\hline			
		\end{tabular}
	\end{center}
    \vspace{-7mm}
	\caption{Abstract Data Organization}
	\vspace{-4mm}
	\label{table:sys_object}
\end{table}

In this section, we define an 
abstract system model in Table~\ref{table:sys_object} 
to ensure clear specification of the protocols in the 
following sections. 
The main purpose of the definition is to provide a mental model to precisely understand the protocol operations, instead of constructing a rigorous formal specification. 
Thus, we omit the unimportant aspects such as
the type of the variables,
and different objects may be presented with a slightly 
different styles, but we believe that the meaning is 
self-explanatory.
Also, the later algorithm description, we mainly specify
the functional aspect of the protocols based on the definition, thus, all objects are considered as global
and can be referred to directly. 

First, we define the total of $F$ files ({\bf File[$F$]}) in the training set, each file can be referred to by its index.
Each file has the static VC index, PC index and O (offset), 
they depend on the chunk size, but can be determined before
the training process and are not changed afterward. 
``H'' indicates the home node of the file: the actual file
data is store in that node. 
``R'' indicates the requester of the file: there can be only
one requester of each file. 
The ``sn'' is the serial number of the file in the global random access trace (updated for each epoch).
The ``addr'' field indicates the starting address of the 
file data: for local read, it is directly returned; 
for remote read, the file data from the address is 
sent to the remote requester. 
The ``consumed'' field indicates whether the file has been
used by the training process. Note that with VC and 
the ``load once'' property (see Introduction), once a file
is loaded into VC, even if it has not been provided
to training, the bit is set. 

The second type of data is all VCs across all nodes ({\bf VC[$M$]}). Each VC has $K$ slots, each slot has 
a valid bit (VFS\_v) and address of the file for this slot 
(addr\_v). For convenience of explanation, we also introduce
VFS for each slot to link it with the global file, so that
in the algorithm we can easily refer to the relevant fields,
but in the actual system, this part is not stored
in the memory
of VC. Each VC has a set of $F/(KM)$ physical chunk that can be mapped
to it, referred to as PCS. 
Finally, each VC also has an ``H'' field indicating
the home node of the VC. 
In our system, each node has the same view of all VCs
but only the local ones can be mapped with the locally 
stored file. A equal set of VCs are reserved
for each remote node to hold the transferred files.

Next, we define the set of physical chunks ({\bf PC[$(F/K)$]}), which is determined
statically before training. 
Each PC has the same number (K) of slots (PFS) as VC.
We also introduce the back pointer VC for convenience.
Similarly, the ``H'' field indicates home of the PC. 
The last part is related to the optional prefetching mechanism 
in the distributed protocol. 
To realize that, we define a prefetch buffer (PFB) for 
each node ({\bf PFB[$N$]}).
The PFB of node-i (PFB[$i$]) maintains the following
information for {\em each} remote node (in total $(N-1)$
but we declare the size to be $N$ for brevity):
(1) a bit map of size $K$ for each entry of the prefetch
window (map[$N$][$K$]);
(2) a pair of VC and O for each prefetched entry
in the prefetch window (p\_VC[$N$][$K$] and P\_O[$N$][$K$]); and
(3) the serial number for the {\em on-demand} file 
request (sn[$N$]).

\subsection{Local Access Protocol}
\label{sec:local_prot}
{
\setlength{\abovecaptionskip}{0pt}  
\setlength{\belowcaptionskip}{0pt}  
\setlength{\textfloatsep}{0pt}  
\begin{algorithm}[!t]
\small
	\SetNoFillComment
	\DontPrintSemicolon
	\caption{Read a file in training dataset}
	\label{algorithm:read_file}
	\SetKwInOut{Input}{Input}
	\SetKwInOut{Output}{Output}
	\SetKwInput{KwRequired}{Required}
	\SetKwFunction{FFindReplacePC}{FindReplacePC}
    \SetKwFunction{FReadLocalFile}{ReadLocalFile}
    \SetKwFunction{FReadRemoteFile}{ReadRemoteFile}
    \SetKwFunction{FReadAndPreFRemoteFile}{ReadAndPreFRemoteFile}
    \SetKwFunction{FReadFile}{ReadFile}
    \SetKwFunction{FCpFileSingle}{CpFileSingle}
    \SetKwFunction{FFillInData}{FillInData}
	\SetKwProg{Fn}{Function}{:}{}
	\SetKw{Break}{break}
	\Input{File to read; Requester}
	\Output{Starting address of a file}

\Fn{\FReadFile{f,R}}{
    \eIf{f.H==R}{
    	\Return \FReadLocalFile{f};
    }{
        addr\_t RFileAddr;\\
        \If{f.VC.VFS\_v[f.O]==true}{
            \Return f.VC.addr\_v[f.O];
        }
        
        \eIf{Prefetch==false}{
            RFileAddr=\FCpFileSingle(\FReadRemoteFile{f,R,f.H});\\

        }{
            RFileAddr=\FFillInData(f,\FReadAndPreFRemoteFile{f,R,f.H});\\
        }
        assert(f.VC.H==f.H); \\
        assert(f.VC.VFS\_v[f.O]==false); \\
        \Return RFileAddr; 
    }
}
\end{algorithm}

\begin{algorithm}[!t]
    \small
	\SetNoFillComment
	\DontPrintSemicolon
	\caption{Read a local file in training dataset}
	\label{algorithm:read_local_file}
	\SetKwInOut{Input}{Input}
	\SetKwInOut{Output}{Output}
	\SetKwInput{KwRequired}{Required}
	\SetKwFunction{FFindReplacePC}{FindReplacePC}
    \SetKwFunction{FReadLocalFile}{ReadLocalFile}
	\SetKwProg{Fn}{Function}{:}{}
	\SetKw{Break}{break}
	\Input{File to read}
	\Output{Starting address of a file}

\Fn{\FReadLocalFile{f}}{
    \eIf{f.VC.VFS\_v[f.O]==true}{
    	f.VC.VFS[f.O].consumed=true; \\
        f.VC.VFS\_v[f.O]=false;  \\
        \Return f.VC.addr\_v[f.O]; \\
    }{
        pc\_t rpc=\FFindReplacePC(f); \\
        \For{(i=0; i<K; i++)} {
            assert(rpc.VC==f.VC); \\
            \If{rpc.PFS[i].consumed==false) \&\& \\
            	 (f.VC.VFS\_v[i]==false} {
                     f.VC.VFS\_v[i]=true; \\
                     f.VC.VFS[i]=rpc.PFS[i]; \\
                     f.VC.addr\_v[i]=rpc.PFS[i].addr; \\
                     f.PC.PFS[i].consumed=true; \\
            	 }
        }
        f.VC.VFS\_v[f.O]=false;\\
        \Return f.VC.VFS[f.O].addr;
    }
}
\end{algorithm}

\begin{algorithm}[!t]
\small
	\SetNoFillComment
	\DontPrintSemicolon
	\caption{Find a proper physical chunk to refill}
	\label{algorithm:file_refill}
	\SetKwInOut{Input}{Input}
	\SetKwInOut{Output}{Output}
	\SetKwInput{KwRequired}{Required}
	\SetKwFunction{FFindReplacePC}{FindReplacePC}
	\SetKwProg{Fn}{Function}{:}{}
	\SetKw{Break}{break}
	\Input{File to read}
	\Output{Selected physical chunk to be fetched}

\Fn{\FFindReplacePC{f}}{
    int usefulRefill=0; \\
    int maxUsefulRefill=-1; \\
    pc\_t candidate\_pc;\\
    \For{(i=0; i<\yuhaosuggest{F/(KM)}; i++)} {
        \If{f.VC.PCS[i].PFS[f.O].consumed==false}{
            \For{(j=0; j<K; j++)}{
                \If{f.VC.VFS\_v[j]==false) \&\&
                    (f.VC.PCS[i].PFS[j].consumed==false}{
                        usefulRefill++;
                    }
                \If{\yuhaosuggest{usefulRefill>maxUsefulRefill}}{
                    candidate\_pc=f.VC.PCS[i]; \\
                    maxUsefulRefill=usefulRefill; \\
                }
            }
        }
    }
    \Return candidate\_pc;
}
\end{algorithm}

\begin{algorithm}[!t]
\small
	\SetNoFillComment
	\DontPrintSemicolon
	\caption{\yuhaosuggest{Read remote file in node-i w/o prefetch}}
	\label{algorithm:remote_no_prefetch}
	\SetKwInOut{Input}{Input}
	\SetKwInOut{Output}{Output}
	\SetKwInput{KwRequired}{Required}
	\SetKwFunction{FFindReplacePC}{FindReplacePC}
    \SetKwFunction{FReadLocalFile}{ReadLocalFile}
    \SetKwFunction{FReadRemoteFile}{ReadRemoteFile}
    \SetKwFunction{FReadFile}{ReadFile}
	\SetKwProg{Fn}{Function}{:}{}
	\SetKw{Break}{break}
	\Input{File; Requester; Home node of the file}
	\Output{Starting address of a file to be sent}

\Fn{\FReadRemoteFile{f,R,H}}{
    \yuhaosuggest{assert(i==f.H);}\\
    addr\_t resp\_file\_addr; \\
    resp\_file\_addr = \FReadLocalFile{f}; \\
    send file data from resp\_file\_addr to node R; \\
}
\end{algorithm}

This and next section specifies the local and remote protocol
with key functions in pseudocode. Our goal is to clearly 
define the operations, rather than seeking the rigorous
function definitions, thus we do not specify the return and argment types, 
and when we define a local variable of certain type, we just 
use the self-explanatory suffix ``\_t'' to indicate the type, e.g., 
``pc\_t'' for a temporary PC.

The interface to read one of the files stored across all nodes is the 
function \texttt{ReadFile} in Algorithm~\ref{algorithm:read_file}.
In a real system, file ID can be passed as the argument, but in our
description, we assume we can access various attributes through {\em f}. 
The second argument indicates the requester's ID, it is used to determine
whether to invoke the local or remote read protocol.
This section considers the local protocol (invoked in line 3), and the 
function \texttt{ReadLocalFile} is specified in Algorithm~\ref{algorithm:read_local_file}.

To read a local file, we check whether slot in the target VC of the file
is valid, if so, we do not check whether the file is the requested one and 
directly return its starting address. Due to the ``read once'' property, 
the slot is marked as invalid and the file is marked as ``consumed''. 
This part captures the key difference of Redox compared to a conventional protocol. 
When the slot in the target VC is not valid, we need to select
a PC to refill into the VC. 
The selection process is specified in an individual function 
\texttt{FindReplacePC} in Algorithm~\ref{algorithm:file_refill}:
to ensure correctness, the refilled PC must at least
contain valid file (not consumed yet)
for the target slot (checked in line 6);
to reduce wasted read, 
we want to maximize the ``useful refill'' (not consumed)
for empty slots of VC (checked in line 8).
After \texttt{FindReplacePC} returns the target PC in Algorithm~\ref{algorithm:read_local_file}
(line 7), we perform the fill for each slot (line 8) when possible, i.e., 
the refill PC's slot should not been consumed and the VC slot should not be valid (line 10 $~\sim$ 11),
and can also validate that the VC of the returned PC and requested f
should be the same (line 9).
For each slot, the refill (line 12 $\sim$ 15)
sets up the valid bit, file, starting address and
the ``consumed'' bit (recall that once a slot enters VC, it is considered as consumed).
Finally, before returning the file of the target slot, we need
to set it to invalid (line 18).

\subsection{Remote Access Protocol Without Prefetch}
\label{sec:remote_prot_nop}

In Algorithm~\ref{algorithm:read_file}, if the requested file's home node
is not the same as the requester, the distributed part of the protocol 
is invoked. It is still possible to satisfy the file request without 
inter-node communication, when the file exist in a VC in the requester node
that holds the remote file data (line 6 $~\sim$ 8). 
In this case, the file data was previously prefetched.
If the prefetch is not enabled, we claim that such scenario should
never happen, simply because between nodes, only the data of the requested {\em individual} 
file---rather than the whole batch of file like from local disk to memory---is transferred.
In another word, there should not been any remote data delivered to the local 
node more than what will be consumed. 

The remote data read protocol without prefetching is specified 
in \texttt{ReadRemoteFile} function in Algorithm~\ref{algorithm:remote_no_prefetch}.
We can see that it essentially leverage the existing local
file read mechanism (line 4) to read the file. 
If the file exist in remote node's VC, the data is directly transferred to the 
requester, otherwise, like we discussed before, the protocol will identify
a PC from the remote node to fill in the target VC.
Here, we can see that there are multiple consumers of the files stored in a node:
one local or $(N-1)$ remote nodes. 
Thus, the PCs will be filled in VCs more quickly, compared to the single-node scenario 
where only the local protocol is invoked. 
When the data of the file is delivered to the requester, 
it is copied into the slot in VC and returned to training process immediately. 
It is implemented in \texttt{CPFileSingle} function and due to its straightforward
functionality, we omit the details. 
Since the file is directly returned, without prefetching, the VCs in a node
for remote data are actually not used, because the slot becomes invalid 
after data return anyway. However, they are indeed used when prefetch is 
enabled, which we discuss next.

\subsection{Remote Access Protocol With Prefetch}
\label{sec:remote_prot_p}

\begin{algorithm}[!t]
\small
	\SetNoFillComment
	\DontPrintSemicolon
	\caption{\yuhaosuggest{Read remote file in node-i w/ prefetch}}
	\label{algorithm:remote_prefetch}
	\SetKwInOut{Input}{Input}
	\SetKwInOut{Output}{Output}
	\SetKwInput{KwRequired}{Required}
	\SetKwFunction{FFindReplacePC}{FindReplacePC}
    \SetKwFunction{FReadLocalFile}{ReadLocalFile}
    \SetKwFunction{FReadRemoteFile}{ReadRemoteFile}
    \SetKwFunction{FReadAndPreFRemoteFile}{ReadAndPreFRemoteFile}
    \SetKwFunction{FReadFile}{ReadFile}
    \SetKwFunction{FGetNextFile}{GetNextFile}
	\SetKwProg{Fn}{Function}{:}{}
	\SetKw{Break}{break}
	\Input{File; Requester; Home node of the file}
	\Output{Send the requested and prefetched file data and map[P] to requester}
    \KwRequired{ \yuhaosuggest{PFB[H].p\_VC[N][P], PFB[H].p\_O[N][P], and PFB[H].sn are initialized as -1s;} PFB[H].map[N][P] is initialized as 0s before training. They are preserved across function calls.}

\Fn{\FReadAndPreFRemoteFile{f,R,H}}{
    \yuhaosuggest{assert(i==f.H);}\\
    addr\_t RespAddr[P]; \\
    file\_t CurFile=f;\\
    int SL=f.sn-PFB[H].sn[R];\\
    
    PFB[H].sn[R]=f.sn; \\
    
    \For{(j=0;j<P;j++)}{
        RespAddr[j]=0;\\
        \yuhaosuggest{PFB[H]}.map[R][j]=0;\\
        \If{j==0}{
            assert(PFB[H].map[R][j+SL]==0);
        }
        \eIf{(j+SL)<P}{
            PFB[H].p\_VC[R][j]=PFB[H].p\_VC[R][j+SL];\\
            PFB[H].p\_O[R][j]=PFB[H].p\_O[R][j+SL];\\
        }{
            PFB[H].p\_VC[R][j]=-1;\\
            PFB[H].p\_O[R][j]=-1;\\
        }
    }

    int next=1;\\
    \For{(j=0;j<P;j++)}{  

            bool C=false; \\
            \For{(s=0;s<j;s++)}{  
                \yuhaosuggest{C|=(CurFile.VC==PFB[H].p\_VC[R][s])\\
                \&\& (CurFile.O==PFB[H].p\_O[R][s]);} \\
            }
            \If{(j==0) ||((CurFile.VC.VFS\_v[CurFile.O]==true) \&\& (!C))}{
                RespAddr[j]=\FReadLocalFile{CurFile}; \\
                PFB[H].p\_VC[R][j]=CurFile.VC; \\
                PFB[H].p\_O[R][j]=CurFile.O; \\
                PFB[H].map[R][j]=1; \\
            }

        \While{File[CurFile.sn+next].R!=R} {
            next++;
        }
        CurFile=File[sn+next];
    } 
    send data of files in RespAddr[P] and PFB[H].map[R][P] to node R;
}

\end{algorithm}

\begin{algorithm}[!t]
\small
	\SetNoFillComment
	\DontPrintSemicolon
	\caption{Requester processes the received on-demand and prefetched data}
	\label{algorithm:fill_remote}
	\SetKwInOut{Input}{Input}
	\SetKwInOut{Output}{Output}
	\SetKwInput{KwRequired}{Required}
	\SetKwFunction{FFindReplacePC}{FindReplacePC}
    \SetKwFunction{FReadLocalFile}{ReadLocalFile}
    \SetKwFunction{FFillInData}{FillInData}
    \SetKwFunction{FReadRemoteFile}{ReadRemoteFile}
    \SetKwFunction{FReadAndPreFRemoteFile}{ReadAndPreFRemoteFile}
    \SetKwFunction{FReadFile}{ReadFile}
    \SetKwFunction{FCpFileMultiple}{CpFileMultiple}
    \SetKwFunction{FFillInData}{FillInData}
    \SetKwFunction{FGetFile}{GetFile}
	\SetKwProg{Fn}{Function}{:}{}
	\SetKw{Break}{break}
	\Input{On-demand requested file; Received data from remote node (multiple file data and map[P]}
	\Output{Starting address of the requested file}

\Fn{\FFillInData{f,RemoteData}}{
    addr\_t FileAddr[P];\\
    int map[P];\\

    FileAddr,map=\FCpFileMultiple{RemoteData};\\
    assert(map[0]==1); \\

    \For{(int i=1;i<P;i++)}{
        \If{map[i]==1}{
            file\_t FillFile=\FGetFile(f.sn,i); \\
            assert(FillFile.VFS\_v[FillFile.O]==0); \\
            FillFile.VFS\_v[FillFile.O]=1; \\
            FillFile.addr\_v[FillFile.O]=FileAddr[i]; \\
        }
    }
    \Return FileAddr[0];
}
\end{algorithm}

Algorithm~\ref{algorithm:remote_prefetch} presents
the protocol to read remote file with prefetching in function \texttt{ReadAndPreFRemoteFile}. 
The arguments of the function includes the file 
requested $f$, its requester R, and the home node H of 
the file. Suppose the code is executed on node i,
i should be the same as H (line 2), since it is the node that
actually stores the file. 
First, we assume that for each epoch, all nodes
are aware of the global random order of file access.
Thus, when an on-demand request of $f$ is received,
with $f$'s serial number (sn), the remote node can 
identify
the next $(P-1)$ file reads in the requester R's 
node as the {\em prefetch window} containing 
the prefetch candidates. 
The requests in prefetch window
are identified in line 38 to line 41, we maintain
a \texttt{CurFile} initialized to be $f$, and then the loop
identifies the next $(P-1)$ requests from node R.

In our system,  
prefetch is not predictive. A natural question is:
if the access trace
is available to all nodes, why not simply fetching all the 
next $(P-1)$ files from node R?
The key reason is: all file data sent from node H
to node R should be successfully inserted into node R's
VC, since these files are 
considered to be ``consumed''.
Note that it is indeed possible that the file cannot be
inserted to the VC at node R, it happens when a prefetched
file is mapped to {\em the same VC and the same offset (O)}.
We call the condition as a {\em conflict}.
To ensure correct prefetch, we need to:
sequentially examine all requests in the prefetch window, starting 
from the on-demand (with the smallest sn), prefetch all 
requests that {\em do not conflict with earlier prefetch
and on-demand requests}. 

The logic is implemented from line 24 to line 33. 
We maintain the (VC,O) pairs
for $P$ requests, i.e., one 
on-demand + $(P-1)$ prefetch candidates, for requester R
in P\_VC[R][0:(P-1)] and P\_O[R][0:(P-1)].
When we send the data of $i$-th file in the
$P$ requests starting from $i=0$ for on-demand request, 
the (VC,O) pairs are recorded in the corresponding entries 
of P\_VC[R][0:(P-1)] and P\_O[R][0:(P-1)].
The i-th bit in {\em map} is set so that the 
receiver side can identify the file. 
If the request is on-demand,
or does not conflict with all earlier requests in the
window {\em and} data is in VC (opportunistic prefetching), 
line 29 to line 32 are executed---obtaining the file, 
recording (VC,O) pair, and updating map. 
Note that \texttt{ReadLocalFile} for on-demand 
request may lead to PC refill.


Moreover, we handle a subtle case when two prefetch windows
{\em overlap}, we need to ensure:
(1) a file should not be prefetched twice; and
(2) the file in the later window should not conflict with
the prefetched data that has not been consumed. 
Consider an example, the first window is
[(a),b,\underline{c},c',\underline{d}], where (a) is the on-demand
request, \underline{c} and \underline{d} are prefetched, and c' conflicts with
c; the second window is [(b),c,c',d,c''], where
(b) is the on-demand request (the second file in 
the first window), and c'' conflict with c and c'.
For the second window,
the correct decision is not to prefetch c and d
since they are prefetched before, and c'' should not
be prefetched since it conflicts with c. 
Essentially, we need to ``remember'' the previously
prefetched files, and use them in the current conflict check.

We can realize this by computing the difference of the current and previous on-demand request's serial number (SL in line 5), if it is smaller than $P$, the two window overlap. 
In this case, (VC,O) pairs of the previous window
can be left shifted by SL, so that previous window's (VC,O) pairs of 
not consumed files are preserved to the current window.
They can prevent prefetching a file twice (a file conflicts
with itself) and files not in the first window.
If SL is equal or greater than $P$, two windows
do not overlap, all current window's (VC,O) pairs should be 
set to (-1,-1), indicating no conflict.
The logic is implemented from line 5 to line 19.
In our example, the (VC,O) pairs after the left-shifting is
(VC,O)[(b),c,c',\_,\_], c'' has conflict with the pair of c, thus cannot 
be prefetched.

Finally, the requester node needs to process the received data, containing
the data of on-demand and a variable number of prefetched files associated with 
a map. It is handled by function \texttt{FillData} in Algorithm~\ref{algorithm:fill_remote}.
The received data is first copied into data buffer with local map by
function \texttt{CpFileMultiple} and the first bit in the map must be 1, since 
this slot is for on-demand request. 
The pointers for the multiple files' data are returned in 
FileAddr[P].
The data of on-demand request (FileAddr[0]) is directly returned (line 14).
For other files with map bit set, the VC is filled, the address
points to the local memory after data copy. 
For the prefetched file, our mechanism 
ensures that the VC slot is empty before the fill (line 9).
The function \texttt{GetFile(f.sn,i} obtains the i-th file after
f in node f.R from the global random sequence. 

}

\subsection{Running Examples}
\label{sec:examples}

To provide the concrete understanding, this section discuss
a few running examples in Figure~\ref{fig:overview}.
First, consider the single-node local access case according to the 
access order in (a). 
For the first four reads, reading the chunks containing the individual files
would lead to waste and random read, since the files are mapped to different 
chunks, as shown in (b).
With Redox protocol, assume the chunks for the four files 
are mapped to the same VC.
After read 1, the whole chunk that contains four
files consecutively stored is batch loaded into memory.
For read 2 and 3, they request the 10th and 15th (index starting from 0)
file, but the target VC of 
the two files' PC is already in memory, and the corresponding slots' are valid, 
so the data of 2nd and 3rd file are returned respectively.
In the original order, these files would have been accessed by read 17 and 7.
With our protocol, at later time, these read will return different files
in the same slots of other PCs mapped to the same VC.
After that, the returned files are marked as consumed. 
At this point at the status in (c), for read 5, although it is also 
mapped to the same VC, but the slot's data has been consumed and is invalid,
at this point, PC refill is needed. 

Next, we consider the distributed protocol with two nodes.
For clarity, they are marked with different colors in (d), 
and each node has its own random sequence, instead of 
a unique global order assumed in earlier protocol description.
The difference is non-essential:
in reality, the sequences are executed
independently. 
The global order in the object system model merely simplifies
the data object specification. 
In (e), we show two on-demand remote requests from node 0:
read 10 and read 15. 
For read 10, when the request is received in node 1, assume
the VC is not filled with the PC.
Node 1 batch reads the PC containing the target file 
to VC, then considers the four future requests from node 0 on node 1
in the prefetch window: read 12, 15, 16, and 17.
In our simple example, all four files' PC have the same (VC=1), so we only 
check offset (O), which are 0, 3, 3, and 1, respectively. 
The (VC,O) of read 10 is (1,3), thus, 
read 15 and 16 conflict with the on-demand read 10, so they 
cannot be prefetched. 
For both read 12 and 17, the files are valid in VC, but read 17 is 
redirected to the file that would have been returned by read 14 in node 1.
In the end, the data of read 10, 12, and 17 are returned to node 0, 
which returns data of read 10 to training process, and inserts 
data for read 12 and 17 to VC. 

Later, when read 15 is sent to node 1, the (VC,O) information of 
the prefetched files (read 17) is left-shifted by two slots.
Note that for read 16, since it was not prefetched, 
its (VC,O) was still (-1,-1), and will not cause any conflict.
The information in the figure
just shows how the conflict is detected. 
At this point, we can see that read 16 again cannot 
be prefetched since it
conflicts with read 15, but the prefetch of read 19 and 20 are 
possible, since they do not conflict with (VC,O) of read 17, and
both VC slots are valid. 
Using our mechanisms, node 1 ``remembers'' 
earlier prefetch of read 17.

\subsection{Implementation Details}

 The core of Redox 
 is implemented in C++. 
The mapping from PCs to VCs 
is implemented using a hash map; the mapping
in the opposite direction is implemented using a combination of a hash map and arrays: each VC maps to an array in the hash map that stores all the PCs it maps to.
The system provides Python interfaces to facilitate its integration with mainstream training frameworks. 
We only replace the underlying module \(data fetcher\) with Redox.
The system encompasses two communication modules: \textit{Inter-Process Communication} (IPC) for facilitating communication between the client and server within the same node, and notably, gRPC-based communication, a protocol specifically designed for efficient and scalable  \textit{remote procedure calls} (RPC) for cross-node data sharing between the servers.

For the distributed protocol, allocating memory space for the all VCs
for remote file would lead to significant waste and affect performance, 
instead, Redox sets a memory limit for the remote VCs. Whenever the requester sends a missed data request to the owner node, it also informs the owner node of the remaining memory space for remote VCs, 
which can be used to determine whether to send the prefetched data. 
	The memory space for the remote VCs
	 is a system hyperparameter, that can be 
	chosen at the value so that the further
	increase beyond it does
	not lead to more prefetched data in memory. 
	In our A10 setup (with 16 GB of memory), allocating approximately 10\% of the memory (1.5 GB) to the remote abstract memory yields the best performance.

\section{Randomness Analysis}
\label{sec:random_diss}

In this section, we discuss Redox protocol's impact on 
randomness. We show that the protocol indeed reduces the 
randomness but the ``remaining randomness'' is still sufficient 
to ensure the training efficiency. In Redox, different VCs
are independent, thus, we just consider one VC of size $K$,
the size of its PCS is $\frac{F}{KM}$, the total number of files in all PCs
mapped to the VC is $\frac{F}{M}$. 

{\bf Definition: } Given a total number of $L$ files, the {\em random
	sequence space} is composed of
the set of all permutations of the $L$ files.
We define {\em randomness} as the size of the random sequence space. 

For a general protocol without file redirection, the randomness of
a sequence of length $L$ is $L!$.  
For a given VC, the length of sequence is $\frac{F}{M}$, thus, 
the randomness is $\frac{F}{M}!$.

{\bf Theorem 1: } File redirection reduces randomness.

{\bf Proof:} After returning
the file in the first slot of PC[0], it is {\em impossible} to return 
any file in the second slot of PC[1] to PC[$\frac{F}{KM}-1$].$\blacksquare$

While the theorem is clear, it does not give the actual randomness with
file redirection. We are interested in understanding whether it becomes
too small that may affect the training efficiency. 
We believe that computing the actual randomness is highly non-trivial
and well beyond the scope of this paper, instead we want to estimate
a lower bound of the randomness. 
With the reduced randomness, 
multiple random sequences are mapped to one random sequence with 
file redirection. 
To estimate the lower bound, we compute the largest number of 
possible sequences in the original space that can be mapped to one sequence
with file redirection.

{\bf Theorem 2: } Given a setting where the chunk size is $K$, 
and size of PCS is $\frac{F}{KM}$, the maximum number of random 
sequences in the original space that can be mapped to one
sequence is $[\frac{F}{KM}!]^K$.

{\bf Proof: } Without loss
of generality, we assume that the first file returned is in the first slot
of PC[0], to return this, 
$\frac{F}{KM}$ choices in the original sequence are possible.
Note that we slightly relax the protocol such that when read a file,
a PC that does not contain the file can be filled into VC.
This change increases the number of mapped original sequence and
reduces lower bound estimation.

Then we try to consider the possible choices for returning
different next files. It is easy to see that, after PC[0] is filled
to VC, it is impossible to return any file in the region of
PC[$1:\frac{F}{KM}-1$][$1:(K-1)$], because a file in this region
is redirected to PC[0][$1:(K-1)$].
For each file in PC[0][$1:(K-1)$], there are $\frac{F}{KM}$ choices.
Now consider the files in the region PC[$1:\frac{F}{KM}-1$][0],
they can be indeed returned, but the number of choices is
$\frac{F}{KM}-1$.
Thus, returning a file in PC[0][$1:(K-1)$] leads to larger number of 
mapped sequences. It is not hard to see that, if we consecutively
return all the $(K-1)$ files in PC[0], each file has $\frac{F}{KM}$ choices.
Based on the above reasoning, it leads to the largest mapped space
for the first $K$ files with size $(\frac{F}{KM})^K$.

The situation for other PCs are similar, except that the choices 
is decreased by one after returning all files of a PC.
Thus, for the second PC, the largest mapped space is $(\frac{F}{KM}-1)^K$;
for the third PC, the largest mapped space is $(\frac{F}{KM}-2)^K$, etc.
Considering all PCs, the largest mapped space is
$(\frac{F}{KM})^K \times (\frac{F}{KM}-1)^K \times (\frac{F}{KM}-2)^K \times ... \times 2^K \times 1^K = (\frac{F}{KM}!)^K$.  $\blacksquare$

Based on theorem 2, the lower bound of randomness with 
file redirection is $(\frac{F}{M}!)/(\frac{F}{KM}!)^K$.
The actual randomness can be larger, since for a file return order
of ``PC-by-PC'' in the proof, the maximum number of original sequences
are mapped to this single order. For other sequence produced by 
file redirection, the mapped space can be smaller, thus randomness
is reduced by a smaller factor. Although $(\frac{F}{KM}!)^K$ is not a 
small number, compared to $(\frac{F}{M}!)$ it is much smaller. 
\yuhao{For example, for $K=64, F=1.28 \times 10^{6}, M=5000$, randomness after file redirection is at least $2.16 \times 10^{88} $. }
We believe it will not affect training efficiency, confirmed by experimental results.

\section{EVALUATION}
\label{sec:evaluation}

\begin{table*}[!htbp]
    \centering
    \caption{Experimental Setup}
    \vspace{-2mm}
    \label{tab:experimentSetup}
    \footnotesize
    \begin{tabular}{>{\centering\arraybackslash}m{0.05\linewidth}
                    >{\centering\arraybackslash}m{0.05\linewidth}
                    >{\centering\arraybackslash}m{0.25\linewidth}
                    >{\centering\arraybackslash}m{0.1\linewidth} 
                    >{\centering\arraybackslash}m{0.1\linewidth} 
                    >{\centering\arraybackslash}m{0.2\linewidth}
        }
        \toprule
        Name & Nodes & GPU/Node & CPU/Node & Memory/Node & Network Bandwidth \\
        \midrule
        A10 & 5 & 1*NVIDIA A10 (24 GB) & 8 Cores & 16 GB & 0.38 GB/s \\
        
        P100 & 3 & 1*NVIDIA P100 (16 GB) & 8 Cores & 60 GB & 0.38 GB/s \\

        A100 & 3 & 4*NVIDIA A100 (40 GB) & 48 Cores & 256 GB & 3 GB/s \\
        \bottomrule
    \end{tabular}
        \vspace{-2mm}

\end{table*}

\subsection{Methodology} 

\textbf{Experimental Setup}. 
Our experiments were conducted in three environments, as summarized in Table \ref{tab:experimentSetup}. 
For A10 and P100 setups, each node is equipped with dual Intel Xeon E5-2682 v4 4-core processors. Training data for these two setups is stored on an Apsara File Storage NAS ~\cite{alibabacloud-nas} with a capacity of 10 PB. In contrast, the A100 setup uses nodes equipped with dual AMD EPYC 7402 24-core processors.  Training data for the A100 nodes is stored on a Lustre file system with a capacity of 2 PB. All the servers run on CentOS Linux 7 (Core) with CUDA Toolkit version 11.2, and PyTorch version 1.7.0 is used for the experiments. The workloads used are shown in Table \ref{tab:workloads}.
    The batch size is set to maximize GPU utilization.

\begin{table*}[!htbp]
  \caption{Workloads}
  \vspace{-3mm}
  \label{tab:workloads}
  \footnotesize
  \centering
  \begin{tabular}{
    >{\centering\arraybackslash}m{0.1\linewidth}
    >{\centering\arraybackslash}m{0.35\linewidth}
    >{\centering\arraybackslash}m{0.15\linewidth}
    >{\centering\arraybackslash}m{0.15\linewidth}
  }
    \toprule
    Task & Dataset & Model & Batch Size \\
    \midrule

    \makecell{Speech \\ Recognition} &
    \makecell[c]{LibriSpeech\textsuperscript{\cite{panayotov2015librispeech}} \\
                 60 GB \,|\, $2.8 \times 10^{5}$ files \,|\, 200 KB avg.} &
    Wav2Vec 2.0\textsuperscript{\cite{baevski2020wav2vec}} &
    64 (A10) \\

    \midrule 

    \multirow{5}{*}{\makecell{Image \\ Classification}} &
    \multirow{3}{=}{\centering\makecell[c]{ImageNet-1k \textsuperscript{\cite{deng2009imagenet}} \\
                 135 GB \,|\, $1.3 \times 10^{6}$ files \,|\, 100 KB avg.}} &
    SqueezeNet\textsuperscript{\cite{iandola2016squeezenet}} &
    512 (A10/P100) \\

    & & MobileNetV3\textsuperscript{\cite{howard2017mobilenets}} & 256 (A10/P100) \\

    & & ResNet50\textsuperscript{\cite{he2016resnet}} & 128 (A10/P100) \\

    \cmidrule(lr){2-4} 

    & \multirow{2}{=}{\centering\makecell[c]{ImageNet-21k\textsuperscript{\cite{deng2009imagenet}} \\
                 1.1 TB \,|\, $1.3 \times 10^{7}$ files \,|\, 100 KB avg.}} &
    DenseNet121\textsuperscript{\cite{huang2017densenet}} &
    256 (A100) \\

    & & VGG16\textsuperscript{\cite{simonyan2014vggnet}} & 256 (A100) \\

    \bottomrule
  \end{tabular}
  \vspace{-5mm}
\end{table*}


\textbf{Baselines}.
(1) PyTorch: uses the native PyTorch DataLoader, relying on the operating system's cache policy for memory management.
(2) No-I/O: based on the native PyTorch DataLoader but replaces the I/O reading by randomly simulating training data generation.
This baseline does not incur any I/O cost and represents the upper bound of any optimization.
(3) CoorDL~\cite{mohan2021coordl}: caches fixed data in memory without memory replacement, representing the state-of-the-art approaches without sacrificing data randomness.

\subsection{Overall Performance}
\label{sec:evaluation-overallPerformance}
We first evaluate training performance in terms of the training time per epoch. 
We report the average training time over five epochs, excluding the first epoch.  All the experiments utilize the maximum available memory whenever possible. In particular, CoorDL can cache 16\% LibriSpeech and 8\% of ImageNet-1k on an A10 node, 35\% of ImageNet-1k on a P100 node and 15\% of ImageNet-21k on an A100 node.

\begin{figure}[ht]
\centering
  \includegraphics[width=0.8\linewidth]{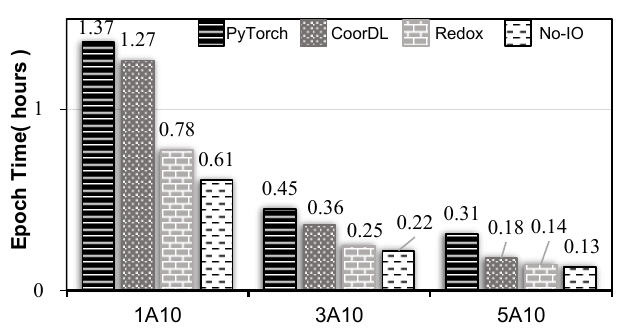}
    \vspace{-4mm}
    \caption{Overall Performance - LibriSpeech}
    \label{fig:overall performance-asr}
      \vspace{-4mm}
\end{figure}

\textbf{LibriSpeech}.  As shown in Figure~\ref{fig:overall performance-asr}, 
when training on one node, Redox outperforms PyTorch, achieving a speedup of 1.77x, and surpasses CoorDL with a speedup of 1.64x. When training on three or five nodes, distributed training enables the exchange of data across nodes. Redox achieves a maximum 2.18x speed improvement over PyTorch and is 1.46x faster than CoorDL during distributed training. Specifically, when training on five A10 GPUs, the performance of Redox is very close to that of No I/O.

\begin{figure*}[htbp]
  \centering
  \begin{subfigure}{0.245\linewidth}
    \centering
    \includegraphics[width=\linewidth]{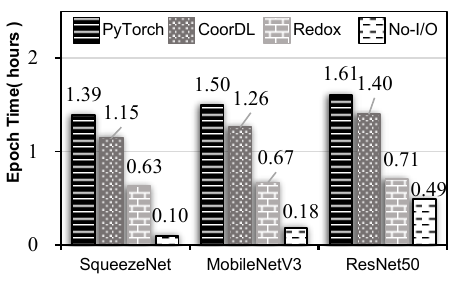}
    \vspace{-5mm}
    \caption{Three A10 Nodes}
    \label{fig:overallPerformance-threeA10Nodes}
  \end{subfigure}%
  \hfill
  \begin{subfigure}{0.245\linewidth}
    \centering
    \includegraphics[width=\linewidth]{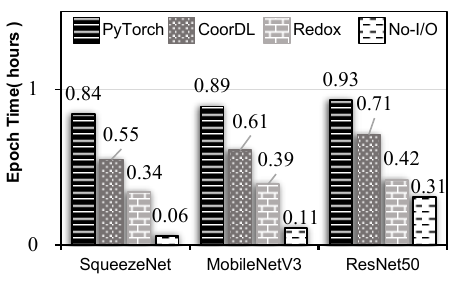}
    \vspace{-5mm}
    \caption{Five A10 Nodes}
    \label{fig:overallPerformance-fiveA10Nodes}
  \end{subfigure}%
  \hfill
  \begin{subfigure}{0.245\linewidth}
    \centering
    \includegraphics[width=\linewidth]{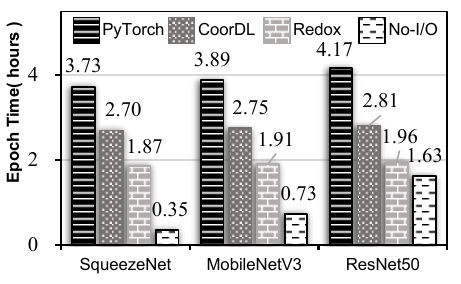}
    \vspace{-5mm}
    \caption{One P100 Node}
    \label{fig:overallPerformance-oneP100Node}
  \end{subfigure}%
  \hfill
  \begin{subfigure}{0.245\linewidth}
    \centering
    \includegraphics[width=\linewidth]{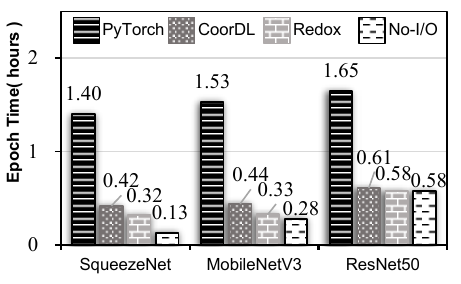}
    \vspace{-5mm}
    \caption{Three P100 Nodes}
    \label{fig:overallPerformance-threeP100Nodes}
  \end{subfigure}%

  \vspace{-3mm}
  \caption{Overall Performance - ImageNet-1k}
  \label{fig:overallPerformance-imagenet-1k}
  \vspace{-5mm}
\end{figure*}

\textbf{ImageNet-1k}. As shown in Figure~\ref{fig:overallPerformance-imagenet-1k}, to evaluate Redox's performance under different memory and hardware conditions, we conducted two sets of experiments. On the A10 setup, we focused on assessing the distributed performance in a scenario with limited memory resources. On the P100 setup, we evaluated performance for both single-node training and distributed training under conditions of ample global memory. When training on three A10 nodes, the global memory can only cache 25\% of ImageNet-1k. As shown in Figure~\ref{fig:overallPerformance-threeA10Nodes}, Redox achieves a maximum 2.26x speed improvement over PyTorch and is 1.96x faster than CoorDL in this case. In Figure~\ref{fig:overallPerformance-fiveA10Nodes}, Redox outperforms PyTorch by up to 2.47x and is 1.69x faster than CoorDL when training on five A10 nodes whose global memory can cache about 40\% data. When training on one P100 node, Redox and CoorDL do not perform data sharing across nodes. In this case, Redox outperforms PyTorch by up to 2.13x, and surpasses CoorDL with a maximum speedup of 1.45x as shown in \ref{fig:overallPerformance-oneP100Node}. The last experiment is conducted on three P100 Nodes whose global memory can cache the entire dataset. As shown in Figure~\ref{fig:overallPerformance-threeP100Nodes}, the performance of Redox is up to 4.57x that of PyTorch. Thanks to the prefetch mechanism, Redox achieves 1.32x better performance compared to CoorDL. Specifically, for GPU-intensive models like ResNet50, the performance of Redox is similar to No-I/O, meaning it does not introduce additional I/O overhead.

\begin{figure}[!htbp]
\centering
  \includegraphics[width=0.8\linewidth]{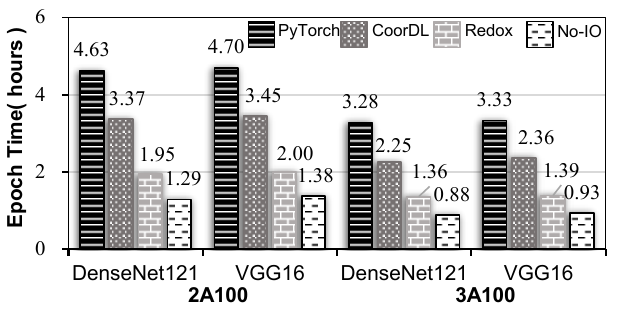}
          \vspace{-5mm}
  \caption{ Overall Performance - ImageNet-21k}
          \vspace{-6mm}
  \label{fig:overallPerformance-A100}
\end{figure}

\textbf{ImageNet-21k}. We validate the performance of two models on two and three A100 nodes for the largest dataset ImageNet-21k.
As shown in Figure~\ref{fig:overallPerformance-A100}, DenseNet121 and VGG16 are evaluated for performance comparison. When training on two nodes, Redox is up to 2.37x faster than PyTorch and up to 1.73x faster than CoorDL. When training on three nodes, Redox is up to 2.41x faster than PyTorch and up to 1.70x faster than CoorDL.

\subsection{Breakdowns Analysis}
\label{sec:eveluation-breakdownsAnalysis}


We compare the performance  by training ResNet50 on ImageNet-1k across three A10 nodes when different optimizations are enabled:
(1) Redox-no-Optimization;
(2) Redox-random-selection without minimizing wasted data read;
(3) Redox-no-prefetching.
We report three key factors: epoch time, memory misses, and remote data requests.

\textbf{Epoch time}. The results are shown 
in Table \ref{tab:breakdownsAnalysis}. 
Even without any optimization,
Redox-no-Optimization considerably 
outperforms the native PyTorch and CoorDL.
Moreover, each optimization does
lead to substantial improvements.

\textbf{Memory misses and remote data requests}.
The prefetching significantly
reduces the remote data accesses and miss
rates. As a side effects, the prefetching
can release VC slots more
frequently, improving the batch data read and 
fill efficiency. Specifically, Redox-no-optimization and Redox-no-prefetching exhibit the same remote data accesses, as both variants do not implement the prefetching strategy. As a result, only the target data is transmitted between nodes, without causing any changes to the remote requests. The results also show 
the importance of refill chunk selection policy, which can further reduce the miss rate. 

\begin{table}[ht]
  \centering
    \footnotesize

  \caption{Breakdowns Analysis}
      \vspace{-3mm}
    \label{tab:breakdownsAnalysis}
    \begin{tabular}{
        >{\centering\arraybackslash}m{0.33\linewidth}
        >{\centering\arraybackslash}m{0.1\linewidth}
        >{\centering\arraybackslash}m{0.18\linewidth}
        >{\centering\arraybackslash}m{0.18\linewidth}
    }
    \toprule
    Frameworks/Variants & Epoch Time & Memory Misses ($\times 10^5$) & Remote Data Requests ($\times 10^5$)\\
    \midrule
     Redox & 0.71 & 1.26 & 0.41 \\
     Redox-random-selection & 0.76 & 1.33 & 0.46 \\
     Redox-no-prefetching & 0.87 & 1.78 & 8.54 \\
     Redox-no-optimization & 0.93 & 1.91 & 8.54 \\
     CoorDL & 1.40 & - & - \\
     PyTorch  & 1.61 & - & - \\
    \bottomrule
    \end{tabular}
      \vspace{-7mm}
\end{table}

\subsection{Remote Abstract Memory Usage}
\label{sec:prefetch-buffer}

\begin{figure}[ht]
\centering
  \vspace{-2mm}
  \includegraphics[width=0.8\linewidth]{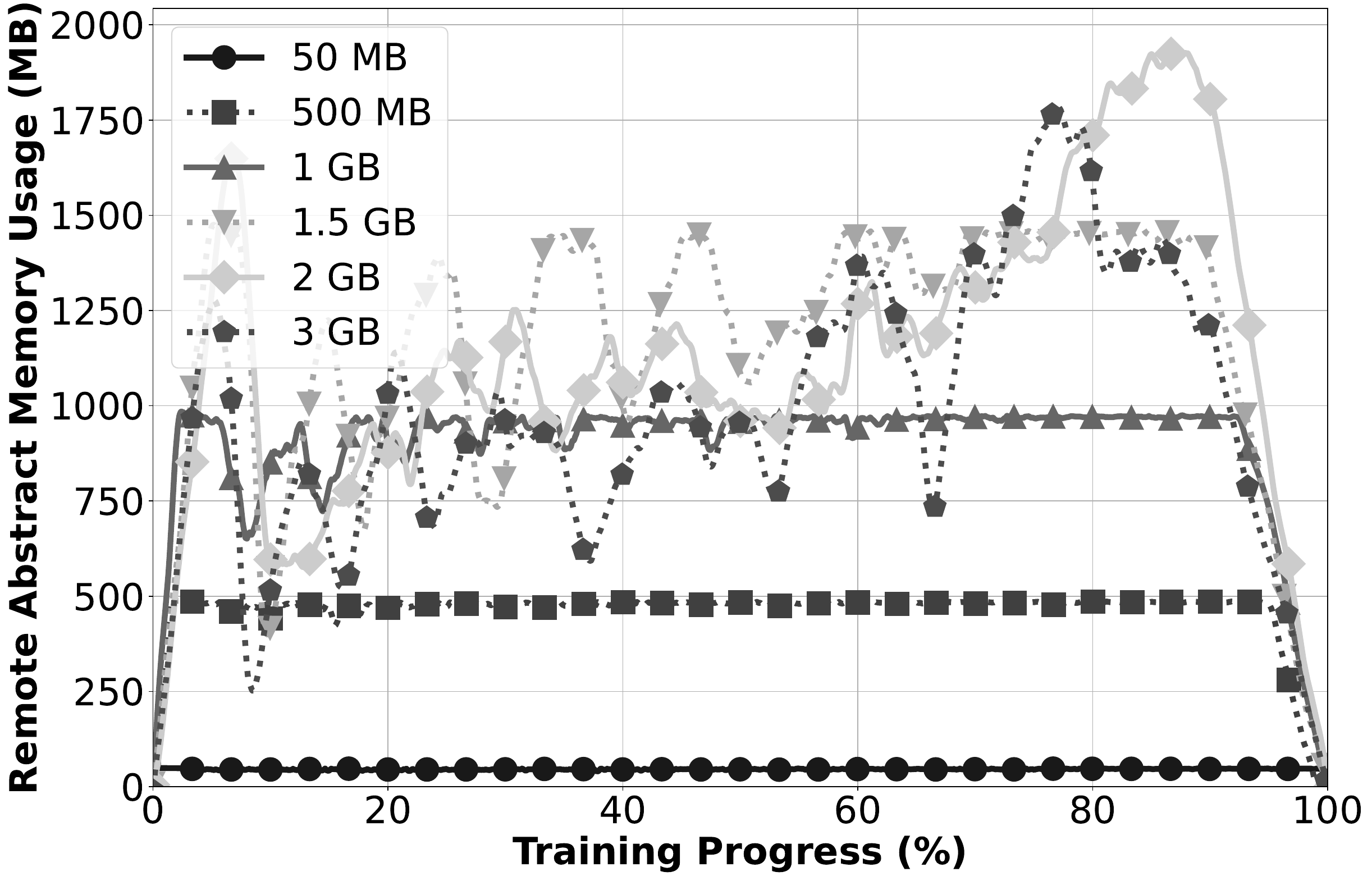}
  \vspace{-2mm}
  \caption{Remote VC Memory Usage Analysis}
  \label{fig:prefetchBufferUsage}
  \vspace{-5mm}
\end{figure}

The experiments are conducted on three A10 nodes using ImageNet-1k as the dataset and SqueezeNet as the model. We manually set the memory limit of remote VCs on each node to 50MB, 500MB, 1GB, 1.5GB, 2GB, and 3GB,
the epoch times (in hours) are 0.77, 0.71, 0.65, 0.63,0.66, 0.68, 
respectively. 
As shown in Figure~\ref{fig:prefetchBufferUsage}, when the memory limit is less than 1.5 GB, increasing this limit leads to a corresponding increase in the memory usage of remote VCs, indicating that more prefetched data are being cached. However, when the limit exceeds 1.5 GB, the usage of memory for remote VCs does not show significant increases. We find that 
the best performance is achieved when the limit is set to 1.5 GB.


\subsection{Chunk Size Sensitivity Study}
\label{sec:eveluation-chunkSize}

\begin{figure}[!htbp]
  \centering
    \vspace{-3mm}
  \begin{subfigure}[b]{0.48\linewidth}
    \centering
    \includegraphics[width=\linewidth]{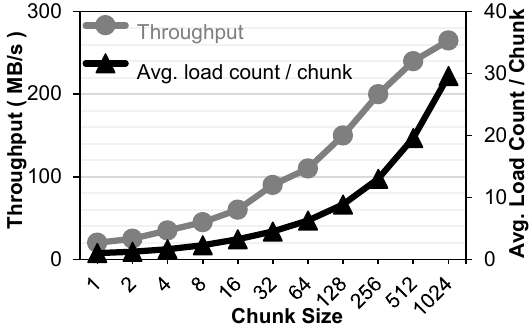}
    \caption{I/O Throughput}
    \label{fig:chunkSize-throughputAndAverageAccessTimes}
  \end{subfigure}
  \hfill
  \begin{subfigure}[b]{0.48\linewidth}
    \centering
    \includegraphics[width=\linewidth]{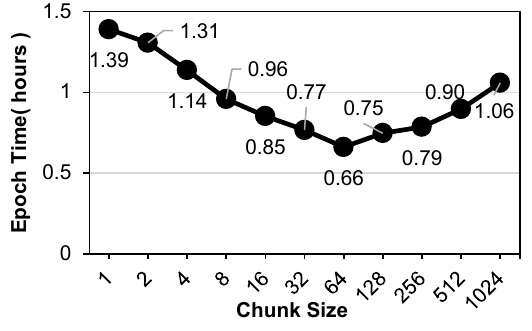}
    \caption{Epoch Time}
    \label{fig:chunkSize-epochTime}
  \end{subfigure}
  \vspace{-3mm}
  \caption{Chunk Size Sensitivity Study}
  \label{fig:chunkSize-SensitivityStudy}
  \vspace{-9mm}
\end{figure}

We analyze the I/O throughput, average number of times each chunk is loaded, and epoch time with chunk sizes varying from 2 to 256, shown in Figure~\ref{fig:chunkSize-throughputAndAverageAccessTimes} and Figure~\ref{fig:chunkSize-epochTime}. 
We compare these results with native PyTorch training with chunk size being 1. 
With chunk size increase, I/O throughput consistently increases, while
the epoch time first decreases up to chunk size 64 and then increases.
It is because when chunk size is too large, a chunk may be loaded 
multiple times and each loading only fills a portion of the data into memory,
leading to more wasted data transfer. Thus, we use chunk size 64 
in our evaluation.  



\subsection{Analysis of Convergence}
\label{sec:convergence}
\begin{figure}[ht]
  \centering
    \vspace{-3mm}
      \includegraphics[width=0.8\linewidth]{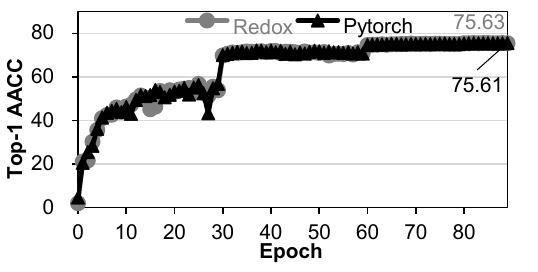}
      \vspace{-5mm}
        \caption{Analysis of Convergence }
        \label{fig:convergence}
        \vspace{-2mm}
\end{figure}

To evaluate the impact on model convergence during training, we train ResNet50 on two P100 nodes using the ImageNet-1k, running for 90 epochs until convergence, an approach that represents the current state-of-the-art evaluation for DNN training convergence. As shown in Figure~\ref{fig:convergence}, throughout the training process, a consistent alignment in Top-1 Accuracy is observed between Redox and PyTorch. Redox achieves its highest accuracy of 75.63 at epoch 83, while PyTorch reaches its peak accuracy of 75.61 at epoch 89. 

To further validate the impact of Redox on convergence, we conduct model convergence experiments with different mappings. For the same dataset, different memory capacities of each node will lead to different mappings. Therefore, we use  the ``stress'' tool in Linux to simulate different memory capacity environments and validate their convergence separately on two P100 nodes. As shown in Table \ref{tab:convergence-differentMapRatios}, regardless of the changes in the memory capacity of each node, neither Top1-ACC and Top5-ACC of Redox are adversely affected compared to native PyTorch. All the experimental results indicate that Redox achieves a similar convergence of model training as vanilla PyTorch does.

\begin{table}[ht]
    \centering
    \vspace{-3mm}
    \caption{Accuracy with Different Memory Size}
    \vspace{-3mm}
    \footnotesize
    \label{tab:convergence-differentMapRatios}
    \begin{tabular}{
        >{\centering\arraybackslash}m{0.4\linewidth}
        >{\centering\arraybackslash}m{0.2\linewidth}
        >{\centering\arraybackslash}m{0.2\linewidth}
    }
        \toprule
         Node Memory Capacity &  Top1-ACC &   Top5-ACC \\
        \midrule
         15 GB & 75.60  & 92.71 \\
        25 GB & 75.65 & 92.70 \\
        35 GB & 75.61 & 92.65 \\
        60 GB & 75.63 & 92.70 \\
        60 GB (PyTorch) & 75.61 & 92.69 \\

        \bottomrule
    \end{tabular}
    \vspace{-8mm}
\end{table}


\section{CONCLUSION}
\label{sec:conclusion}

This paper proposes Redox, a training data management system designed to achieve high I/O efficiency.
We reveal a new observation named {\em file redirection}: 
for model training, when  training data in one file is requested, 
the system has the flexibility
to return the data of another file. 
Based on this property, 
we propose efficient local and distributed file read protocol
that both minimizes the wasted data read
and enables opportunistic prefetch from remote node. 
Experimental results indicate that Redox significantly accelerates data fetching in training, achieving up to a 4.57x improvement in end-to-end training compared to PyTorch.

{\footnotesize \bibliographystyle{acm}
\bibliography{sample}}


\end{document}